\documentclass[12pt]{article}
\usepackage{amsmath,amsfonts,setspace,indentfirst,bbold}
\usepackage[english]{babel}
\usepackage{epsfig}
\usepackage{slashed}

\def\be{\begin{equation}}
\def\ee{\end{equation}}
\def\ben{\begin{equation*}}
\def\een{\end{equation*}}
\def\ba{\begin{array}}
\def\ea{\end{array}}
\def\bn{\begin{align}}
\def\en{\end{align}}
\def\bnn{\begin{eqnarray*}}
\def\enn{\end{eqnarray*}}
\def\bsub{\begin{subequations}}
\def\esub{\end{subequations}}
\def\p{{\partial}}

\def\bs{\begin{subequations}}
\def\es{\end{subequations}}
\def\Cl{{\mathit{Cl}}}
\def\Spin{{\mathit{Spin}}}
\def\G{{\Gamma}}
\def\g{{\gamma}}
\def\k{{\kappa}}
\def\e{{\epsilon}}
\def\f{{\phi}}
\def\m{{\mu}}
\def\n{{\nu}}
\def\d{{\delta}}
\def\r{{\rho}}
\def\s{{\sigma}}

\def\l{{\lambda}}
\def\eh{{\hat\epsilon}}
\def\a{{\alpha}}
\def\b{{\beta}}
\def\S{{\Sigma}}

\numberwithin{equation}{section}

\begin{document}

\begin{titlepage}

\begin{center}

\hfill{QMUL-PH-09-28} \\

\vskip 2.5cm

{{\Large \bf Exploring Fermionic T-duality}} \\

\vskip 1.25cm Ilya Bakhmatov\footnote{i.bakhmatov@qmul.ac.uk} and  David S. Berman\footnote{D.S.Berman@qmul.ac.uk}
\\
{\vskip 0.5cm
Queen Mary University of London,\\
Department of Physics,\\
Mile End Road, London, E1 4NS, England\\
}

\vspace{30pt}
\begin{abstract}
\baselineskip=18pt\
The fermionic T-duality transformation developed by Berkovits and
Maldacena is applied to D-brane and the pp-wave solutions of type
IIB supergravity. The pp-wave is found to be self-dual under the combination
of dualities. We explore the consequences of applying the
transformation and discuss various properties of the new transformed solutions.
\end{abstract}
\end{center}
\end{titlepage}

\tableofcontents

\section{Introduction}

T-duality is one of the remarkable features of string
theory \cite{eliezer}. It is a map between different string backgrounds that
leaves the partition function of the string sigma model invariant \cite{Buscher}. From
the point of view of the world sheet it is an abelian two-dimensional
S-duality. From the spacetime viewpoint it is somewhat mysterious since it
provides an equivalence between completely different geometries. A key
application of the duality is to use this
symmetry as a solution generating mechanism in supergravity
\cite{Berg} where one begins with a particular solution and then
through application of the T-duality rules produces a new set of
solutions. This technique has proved particularly useful in constructing
solutions deformed by NS flux such as for the gravity duals of
noncommutative theories \cite{maldanc}, beta deformed Yang-Mills \cite{lunin} and so-called dipole
deformed theories \cite{dipole} (similar techniques have also been used for
deformation of M-theory geometries\cite{laura}).

T-duality is also crucial in establishing the connection between the different branes of
type II string theory and has been a central pillar in string duality
for many years. It is surprising then that it was not until 2008 that fermionic
T-duality was developed \cite{BerkMald,Beisert:2008iq}. Usual bosonic T-duality
relies on using an isometry of the background to generate the
T-duality transformation. Fermionic T-duality can be viewed as extending this
idea to isometries of the fermionic directions in superspace. From the viewpoint of
supergravity component fields these are just the supersymmetry transformations, thus
instead of using isometries to generate the T-duality transformations
one uses the supersymmetries. The details of the transformation will be reviewed later in the
paper.

Let us note various aspects of this transformation. Firstly,
it is not a full symmetry of string theory like bosonic T-duality since it is
broken at one loop in $g_s$. (This is because of the presence of
fermionic zero modes in the path integral beyond tree level that make
the path integral vanish. It is interesting to consider if one could
extend the duality beyond tree level by soaking up these zero modes
and making sense of such a path integral including the fermionic insertion).

Secondly, apart from a shift in the dilaton the NSNS sector of the
theory is left invariant. Fermionic T-duality is a transformation primarily of the RR
fields. (This really explains the delay in the study of fermionic
T-duality; deriving the RR transformations in bosonic T-duality from
the string world sheet was only done recently and required using the Berkovits
formulation \cite{policastro}).

Thirdly, because of the requirement that we deal with commuting
supersymmetries (just as one deals with commuting isometries in
ordinary T-duality) it is necessary that we deal with complexified Killing
spinors and in turn complexified RR-fluxes. Thus the transformed
background will be a solution of complexified supergravity. One open and indeed
crucial question is to determine when these transformations map back
to a real supergravity solution. In fact, one need not map directly
to a purely real solution since if there exists a time-like isometry
(which is almost certain for a supersymmetric solution) then one can
do bosonic T-duality in the timelike direction \cite{hulltime}. This
transformation has the effect (amongst other things) of multiplying
the RR-fluxes by an imaginary unit.
Thus it can make purely imaginary fluxes real. This was precisely the
case for the fermionic dual of $AdS_5\times S^5$ described by Berkovits
and Maldacena where after eight fermionic T-dualities there remained some
imaginary RR flux. This was then made real by application of timelike T-duality.

In any case, perhaps we should be interested in complexified supergravity in its own right. In
quantum field theory (such as Yang-Mills) there has been a great deal of progress made by
complexifying the theory and then using the power of complex
analysis. This was the origin of the S-matrix programme which has now
seen something of a revival \cite{gab}  with recent works on amplitude
physics again relying on an implicit complexification of the theory to
achieve results. In fact, the motivation  for
studying fermionic T-duality \cite{BerkMald,Beisert:2008iq} was to derive the duality between certain amplitudes and Wilson lines in
Yang-Mills theory{\footnote{DSB is gratefull to various participants
    of the FPUK meeting in Cambridge for discussions on this
    issue.}}. Whether we can learn really more about string theory per se
through complexification of backgrounds has yet to be seen but ideas
along these lines have appeared before (see for example the discussion
in \cite{hulltime}).

Related work on fermionic T-duality has appeared in \cite{oz,grassi,Chen}.

\section{Fermionic T-duality}

Here we review in detail the fermionic T-duality transformation
procedure derived in \cite{BerkMald}. The type II supersymmetry transformations are
parameterized by the Killing spinors of the solution. These Killing
spinors will determine the transformed solution as follows.

Take, $\e$, a Killing spinor that parameterizes an unbroken
supersymmetry.
It is a Majorana-Weyl spinor of $(1+9)$-dimensional spacetime,
that is, real with sixteen components. Since type II supergravity is an
$\mathcal{N}=2$ theory, there is also another Killing spinor, which is
denoted by $\eh$. A pair $e=(\e,\eh)$ generates one supersymmetry
transformation. However, the two spinors within this
pair are not independent -- they are related by the Killing spinor
equations (see section \ref{D1susy} below), and furthermore by the constraint
\be
\label{BM:constr}
\e\g_\m\e + \eh\g_\m\eh = 0
\ee
for all $\m\in \{0,\ldots,9\}$. Here $\g_\m$ are blocks comprising the 10-dimensional
gamma-matrices in the Weyl representation (see appendix \ref{gamma}).
This constraint arises from insisting
that the supersymmetry with which one carries out the T-duality
transfromation is commuting just as one requires that the Killing
vectors in bosonic T-duality commute. Since $\g_0$ is a unit matrix,
the above relation cannot hold for real spinors, and they must be
artificially complexified. This is a characteristic property of
fermionic T-duality, which then leads to complex RR fluxes after the transformation.

After the choice of the Killing spinors satisfying (\ref{BM:constr}) has been made,
one calculates an auxilliary scalar field $C$ defined by the following differential equation:
\be
\p_\m C = i \e\g_\m\e - i \eh\g_\m\eh.
\ee
By using the constraint (\ref{BM:constr}) we can simplify this to be:
\be
\label{BM:C}
\p_\m C = 2i \e\g_\m\e.
\ee
The transformation of the dilaton is given by
\be
\label{BM:dil}
\f' = \f + \frac12 \log C,
\ee
and the RR forms transformation can be written succinctly in terms of the bispinor $F^{\a\b}$:
\be
\label{BM:RR}
\frac{i}{16}e^{\f'}F' = \frac{i}{16}e^{\f}F - \frac{\e\otimes\eh}{C}.
\ee
The RR field strength bispinor incorporates all RR forms of IIB supergravity:
\be
\label{BM:bispinor}
F^{\a\b} = (\g^\m)^{\a\b} F_\m + \frac{1}{3!} (\g^{\m_1 \m_2 \m_3})^{\a\b} F_{\m_1 \m_2 \m_3} + \frac12 \frac{1}{5!} (\g^{\m_1 \hdots \m_5})^{\a\b} F_{\m_1 \hdots \m_5}.
\ee
We have a factor of $+16$ as compared to $-4$ of \cite{BerkMald}
in the transformation law (\ref{BM:RR}) because of a different
normalisation of RR fields, which is implied by the action (\ref{action}) that we use.
In fact, the formula (\ref{BM:bispinor}) is only correct for
backgrounds with trivial NS two-form (which is the case for the D-brane and pp-wave
backgrounds). If there is a nontrivial $B$-field, then instead of just the RR field strengths one should use the
modified RR field strengths that are invariant under the supergravity gauge transformations as given in
equation (\ref{mod}). This correction is beyond the first order in component fields and
thus was omitted from the original derivation{\footnote{We are grateful
  to Giuseppe Policastro and Nathan Berkovits for clarifying this point.}}.

In the case when the fermionic T-duality is performed with respect to several supersymmetries, parameterized by the Killing spinors $e_i=(\e_i, \eh_i), i\in\{1,\ldots,n\}$, the formulae (\ref{BM:C}), (\ref{BM:dil}), and (\ref{BM:RR}) are generalized to
\bsub
\label{mult-T-d}
\be
\label{C-mult}
\p_\m C_{ij} = 2i \e_i \g_\m \e_j,
\ee
\be
\label{dil-mult}
\f' = \f + \frac12 \sum_{i=1}^n (\log C)_{ii},
\ee
\be
\label{RR-mult}
\frac{i}{16} e^{\f'} F' = \frac{i}{16} e^\f F - \sum_{i,j=1}^n(\e_i \otimes \eh_j)\, (C^{-1})_{ij}.
\ee
\esub
The set of the Killing spinors must obey
\be
\label{abel}
\e_i\g_\m\e_j + \eh_i\g_\m\eh_j = 0
\ee
for all $i,j\in\{1,\ldots,n\}$.

In summary, the recipe to perform fermionic T-duality on a given solution is as follows:
\begin{enumerate}
\item Find the Killing spinors of the solution. In IIB supergravity
these are represented by pairs $e=(\e,\eh)$ of 16-component real spinors of the same chirality.
\item Choose a complex linear combination of the Killing spinors $e'=(\e',\eh')$
      that satisfies the condition (\ref{BM:constr}). These Killing
      spinors describe the supersymmetry that we a dualising with
      respect to.
\item Calculate $C$ from (\ref{BM:C}). To do this consistently, one
      should work in world indices (i.e. one should integrate
$\p_{\underline \m} C = 2i \e' \left(e^\n_{\underline \m} \g_\n\right) \e'$,
where $e^\n_{\underline \m}$ is the vielbein, and world indices are underlined to distinguish them from flat ones).
\item If there are any RR fields in the original background,
      substitute them into (\ref{BM:bispinor}) to calculate the matrix $F^{\a\b}$.
\item Use $F^{\a\b}, \e^\a, \eh^\b$, and $C$ to calculate the transformed RR background $F'^{\a\b}$ via (\ref{BM:RR}).
\item Use (\ref{BM:bispinor}) again, this time to find the contributions of $F_1, F_3$, and $F_5$ to $F'^{\a\b}$ separately.
\item Check that the transformed background is a solution to the field equaitons.
\end{enumerate}

Since the above recipe of doing fermionic T-duality involves a great
deal of 16 by 16 matrix manipulations, it is the easiest to implement
it using a simple {\it Mathematica} program to perform steps 2,4,5,
and 6. (A copy of the program is available via an email to the authors).
The only nontrivial step in such program is number 6, where one starts with a 16 by 16
matrix $F'$, and one needs to find the corresponding 1-, 3-, and
5-form components. This calculation is done by separating the matrices
in equation (\ref{BM:bispinor}) into their symmetric and
antisymmetric parts. On the left-hand side of the equation
we have a matrix $F'$, which has been calculated from (\ref{BM:RR}).
This should be split into symmetric and antisymmetric parts by brute
force. As to the right-hand side of (\ref{BM:bispinor}),
it is naturally separated into symmetric and antisymemtric parts. Namely,
a single $\g$-matrix is symmetric, as well as a product of five $\g$-matrices,
whereas a triple product is antisymmetric. This can be verified explicitly 
by using the matrix representation given in appendix \ref{gamma}.

\section{Fermionic T-duals of the D1-brane}
\label{D1susy}

We begin with the D1-brane solution in IIB. This background has
nonzero dilaton, metric and RR 2-form potential.
These are given by the following \cite{HorStrom}:
\be
\label{D1:phi}
e^{2\f} = 1+\frac{Q}{(\d_{mn} x^m x^n)^6};
\ee
\be
\label{D1:G}
g_{\m\n} = (e^{-\f} \eta_{ij},e^{\f} \d_{mn}),
\ee
\be
g^{\m\n} = (e^{\f} \eta^{ij},e^{-\f} \d^{mn}),
\ee
\be
\sqrt{|g|} = e^{3\f},
\ee
\be
\label{D1:C_2}
(C_2)_{01} = e^{-2\f}-1;\quad (F_3)_{01m} = -2 e^{-2\f} \p_m \f,
\ee
and all the other fields ($B_2, C_0, C_4$) are zero. The notation is
$$
\eta_{ij} = \mathrm{diag} (-1,1),\qquad i,j \in \{0,1\},
$$
$$
\delta_{mn} = \mathrm{diag} (1,1,1,1,1,1,1,1),\qquad m,n \in \{2,\ldots,9\}.
$$
All components of $C_2$ and $F_3$, other than specified in
(\ref{D1:C_2}), are zero.
The indices in (\ref{D1:C_2}) are world indices.

The form of the transformed solution depends on the choice of the
Killing spinor used for the transformation. So a few words about D-brane Killing spinors are in order.

Dp-branes are invariant under the supersymmetry transformations
parameterized by the spinors that satisfy the following projection condition:
\be
(1\pm\G^{0\ldots p} \mathcal{O}) \varepsilon = 0,
\ee
where $\mathcal{O}$ is an operator that depends on the supergravity
type and on the dimensionality of the brane in question.
Thus a generic type II D-brane in ten dimensions has
sixteen unbroken supersymmetries generated by the Killing spinors that satisfy the above constraint.

Confining our attention to the case of D1-brane we have
\be
\varepsilon = \left(\begin{array}{c} \e \\ \eh \end{array}\right),
\ee
where $\e$ and $\eh$ are the two chiral Majorana-Weyl spinors that are
the supersymmetry parameters of type IIB supergravity.
This is written in the two-component formalism, so that $e$
is just a two-component column vector, not a 32-component 10d spinor.
The operator $\mathcal{O}$ is given by the Pauli matrix $\s_1$, so that the Killing spinor constraint takes the form
\be
(1\pm\G^{01} \s_1) \varepsilon = \left(\begin{array}{c} \e \\ \eh \end{array}\right) \pm \left(\begin{array}{c} \G^{01}\eh \\ \G^{01}\e \end{array}\right) = 0,
\ee

Taking the minus sign for definitness we see that, for example, we can
take $\e$ to be arbitrary 16-component
MW spinor, in which case $\eh = \G^{01} \e$.

Technically, the above algebraic constraint on the Killing spinor can
be thought of as arising from the requirement that the supersymmetry
variation of dilatino vanishes. There is also an analogous requirement
for gravitino, which is the second fermionic doublet in type II
supergravity. Since the variation of gravitino contains derivatives of
the supersymmetry parameter, this second constraint leads to a
differential equation on $\varepsilon$. Solving this equation
introduces coordinate dependence into the Killing spinor (note that so
far our $\e$ and $\eh$ were constant). Thus, it turns out that
\be
\e = e^{-\frac{\f}{4}} \e_0
\ee
for an arbitrary constant $\e_0$, and $\eh = \G^{01} \e$, as before.
The function $e^\f$ has been defined in (\ref{D1:phi}).

Using explicit realisation of the gamma-matrices, we see that corresponding to an arbitrary
\be
\e = \left( \e_1,\; \e_2,\; \ldots \e_{16} \right)^T,
\ee
is
\be
\eh = \left( \e_{16},\; -\e_{15},\; -\e_{14},\; \e_{13},\; -\e_{12},\; \e_{11},\; \e_{10},\; -\e_9,\; -\e_8,\; \e_7,\; \e_6,\; -\e_5,\; \e_4,\; -\e_3,\; -\e_2,\; \e_1 \right)^T,
\ee
where the factors of $e^{-\frac{\f}{4}}$
have been omitted for simplicity ($^T$ means transpose, so that $\e$
and $\eh$ are columns). Setting all $\e_i$ but $\e_1$ to zero,
we get the first basis element (which we call $e_1$). It includes both
$\e$ and $\eh$. Repeating this process for all of the sixteen
parameters, we end up with the sixteen basis elements ${e_i}$.

The next step in our programme is to pick a particular linear
combination of the Killing spinors,
so that it satisfies the condition (\ref{BM:constr}).
As mentioned earlier, this constraint cannot be satisfied by real
Killing spinors.
We consider the simplest possible linear combinations, i.e. those of the form
\be
\label{comb}
e' = e_a + i e_b;\, a,b\in\{1,\ldots,16\}.
\ee
Using the explicit form of gamma-matrices it is easy to check that (\ref{BM:constr}) is satisfied by {\it any} such combination, apart from those of the form $e_a + i e_{17-a}$.

The result of the fermionic T-duality transformation can be of two types depending on the values of $a$ and $b$ in (\ref{comb}):
\begin{itemize}
\item If ($a\leq 8$ and $b\leq 8$), or ($a\geq 9$ and $b\geq 9$), then the result is of the `simple' type. This is characterized by vanishing $\e\g_\m\e$, which means that $C$ in (\ref{BM:dil}) and (\ref{BM:RR}) is just a constant. The dilaton is shifted by a constant, the RR field components that were present in the original background (\ref{D1:C_2}) are multiplied by a constant, and several new components of $F_3$ and $F_5$ emerge.
\item If ($a\leq 8$ and $b\geq 9$), or ($a\geq 9$ and $b\leq 8$), then the result is of the `complicated' type. Despite $\e\g_\m\e + \eh\g_\m\eh$ is still zero, as required by (\ref{BM:constr}), $\e\g_\m\e$ is nonzero in this case. This means that $C$ is not a constant (in our examples $C$ will be a linear complex-valued function of the coordinates transverse to the brane, see below). The dilaton is shifted by a logarithm of this function, the RR fields are scaled by a power of it, and some new components of $F_3$ and $F_5$ appear again, but also the components that were present in the original solution (\ref{D1:C_2}) get additive terms.
\end{itemize}

Let us give some explicit examples. As a representative of the
`simple' group of transformed D1-branes we will consider the result of
the duality with a Killing spinor parameter $e_1 + i e_2$. For a
`complicated' class of backgrounds we will use $e_1 + i e_9$. In both
cases we have the same metric (\ref{D1:G}) and $B$-field (zero), as in
the original D1-brane solution -- this is a general property of
fermionic T-duality. In the particular case of D1-brane and for
Killing spinor combinations of the form (\ref{comb}) it turns out that
RR scalar is also the same before and after the transformation
(zero). Shown below are transformed dilaton and the new RR fields.

\subsection{`Simple' case}
\label{easy}

Taking a Killing spinor parameter of the transformation to be
\be
e_1 + i e_2 =
\left\{
\begin{aligned}
&\{1,i,0,0,0,0,0,0,0,0,0,0,0,0,0,0\}\\
&\{0,0,0,0,0,0,0,0,0,0,0,0,0,0,-i,1\}
\end{aligned}
\right\},
\ee
we get from (\ref{BM:C})
\be
\p_\m C = 0,\: \forall \m \quad\Rightarrow\quad C = \mathrm{const}.
\ee
Thus, the dilaton dependence after the duality is
\be
\label{phinew}
e^{2\f'} = C\left(1+\frac{Q}{(\d_{mn} x^m x^n)^6}\right).
\ee
The R-R 3-form has the components (world indices are used everywhere)
\be
\label{F_3new_}
(F_3)_{01m} = -2 C^{-1/2} e^{-2\f} \p_m \f
\ee
(compare to (\ref{D1:C_2})) and eight new constant components
\be
\label{F_3new}
\begin{array}{llll}
F_{236} = i, & F_{268} = -1, & F_{356} = 1, & F_{568} = -i, \\
F_{237} = 1, & F_{278} = i, & F_{357} = -i, & F_{578} = -1.
\end{array}
\ee

There also appear 16 constant components of the 5-form:
\begin{subequations}
\label{F_5new}
\be
\label{F_5:1}
\begin{array}{llll}
F_{02369} = -i, & F_{02689} = 1, & F_{03569} = -1, & F_{05689} = i, \\
F_{02379} = -1, & F_{02789} = -i, & F_{03579} = i, & F_{05789} = 1, \\
\end{array}
\ee
\be
\label{F_5:2}
\begin{array}{llll}
F_{14578} = -i, & F_{13457} = -1, & F_{12478} = 1, & F_{12347} = i, \\
F_{14568} = 1, & F_{13456} = -i, & F_{12468} = i, & F_{12346} = -1.
\end{array}
\ee
\end{subequations}
Note that the indices in (\ref{F_5:1}) result from appending 0 and 9
to the indices of the 3-form components in (\ref{F_3new}).
The components in (\ref{F_5:2}) are required by the self-duality.
All the values given in (\ref{F_3new}) and (\ref{F_5new}) must be additionally multiplied by $2 C^{-3/2}$.

\subsection{`Complicated' case}
\label{hard}

As an example of this type of a transformed background let's take the following linear combination of Killing spinors:
\be
e_1 + i e_9 =
\left\{
\begin{aligned}
&\{1,0,0,0,0,0,0,0,i,0,0,0,0,0,0,0\}\\
&\{0,0,0,0,0,0,0,-i,0,0,0,0,0,0,0,1\}
\end{aligned}
\right\},
\ee
from which it follows that
\be
\label{C}
\p_{0,\ldots,7} C = 0,\, \p_8 C = -4,\, \p_9 C = 4i \quad\Rightarrow\quad C = 4i (x^9 + i x^8).
\ee
We see that the dilaton is now complex-valued:
\be
e^{2\f'} = C e^{2\f} = 4i (x^9 + i x^8)\,\left(1+\frac{Q}{(\d_{mn} x^m x^n)^6}\right).
\ee
The RR fields transform similarly to the simple case with one
important difference: of the eight
newly appearing components of the 3-form only six have truly new indices:
\be
\label{F_3new1}
\begin{array}{lll}
F_{278} = i, & F_{348} = -i, & F_{568} = -i, \\
F_{279} = 1, & F_{349} =-1, & F_{569} = -1,
\end{array}
\ee
whereas the lacking two appear as additive contributions to the $(01m)$ components that were present before the transformation:
\begin{align}
\label{F_3new2}
&F_{012} = -2 C^{-1/2} e^{-2\f} \p_2 \f, \quad\ldots\quad, F_{017} = -2 C^{-1/2} e^{-2\f} \p_7 \f, \nonumber \\
&F_{018} = -2 C^{-1/2} e^{-2\f} \left( \p_8 \f - C^{-1} \right),\quad F_{019} = -2 C^{-1/2} e^{-2\f} \left( \p_9 \f + i C^{-1} \right).
\end{align}
Again there are sixteen components of self-dual 5-form field strength. These components, as well as those of the 3-form in (\ref{F_3new1}), should be multiplied by $2 C^{-3/2}$:
\be
\label{F_5new__}
\begin{array}{llll}
F_{02368} = 1, & F_{02458} = 1, & F_{03578} = -1, & F_{04678} = 1, \\
F_{02369} = -i, & F_{02459} = -i, & F_{03579} = i, & F_{04679} = -i, \\
F_{14579} = -1, & F_{13679} = -1, & F_{12469} = 1, & F_{12359} = -1, \\
F_{14578} = -i, & F_{13678} = -i, & F_{12468} = i, & F_{12358} = -i.
\end{array}
\ee

\subsection{Solution checking the fermionic T-dual}

We have verifed that the transformed backgrounds are indeed solutions
to type IIB supergravity equations of motion.

In the so called, `simple' case, all the equations are trival apart
from the Einstein equation (\ref{_G}) which is satisfied by the transformed solution because the RR fields' energy-momentum tensors change trivially under the transformation -- being quadratic in RR field strengths that scale as $C^{-1/2}$ (\ref{F_3new_}), they simply get multiplied by $C^{-1}$, which is cancelled by the transformation of the dilaton:
\be
\frac{e^{2\f'}}{2} \left[ {T'}_{\m\n}^{(1)} + {T'}_{\m\n}^{(3)} + \frac12 {T'}_{\m\n}^{(5)} \right] =  \frac{C e^{2\f}}{2} \left[\frac{1}{C} T_{\m\n}^{(1)} + \frac{1}{C} T_{\m\n}^{(3)} + \frac{1}{C} \frac12 T_{\m\n}^{(5)} \right],
\ee
so that the right-hand side of (\ref{_G}) does not change (the left-hand side does not change trivially because the dilaton is shifted by a constant and because the curvature is not affected).

An interesting question, however, is how it so happens that the new
components of the 3- and 5-form do not contribute to the
energy-momentum tensor. The reason is an accurate balance of real and
imaginary units, scattered around (\ref{F_3new}) and
(\ref{F_5new}). 

In the so called, `complicated' case, the auxilliary field $C$ in the
transformation is no longer constant. As a result the function $C = 4i
(x^9 + i x^8)$ (\ref{C}) enters into the expressions for the
transformed fields and the verification of most equations is nontrivial.

To gain a flavour of the cancellations involved we will give an
example of solving the dilaton field equation (\ref{_phi}). Using
\be
\f' = \f + \frac12 \log C,
\ee
we calculate
\be
\nabla^2 \f' = \frac{1}{\sqrt{|g|}} \p_m \left( \sqrt{|g|} g^{mn} \p_n \f' \right) = -\frac{e^{-\f}}{2C^2} \d^{mn} \left(\p_m C \p_n C - 2C \p_m \f \p_n C \right),
\ee
\be
(\p\f')^2 = e^{-\f} \d^{mn} \left( \p_m \f \p_n \f + \frac{1}{C} \p_m \f \p_n C + \frac{1}{4C^2} \p_m C \p_n C \right),
\ee
where we have taken into account that for the dilaton in the D1-brane background
\be
\d^{mn} \left( \p_m\p_n\f + 2 \p_m\f\p_n\f \right) \equiv 0,
\ee
and that the second derivatives of $C$ vanish.

For the function $C = 4i(x^9 + i x^8)$ we get
\begin{align}
&\d^{mn} \p_m C \p_n C = (\p_8 C)^2 + (\p_9 C)^2 = 0,\\
&\d^{mn} \p_m \f\, \p_n C = -4 (\p_8 \f - i\, \p_9 \f),
\end{align}
and substituting this into the dilaton field equation (\ref{_phi}) yields
\begin{align}
R + 4 \nabla^2 \f' - 4(\p\f')^2 &= -5e^{-\f} \d^{mn} \p_m \p_n \f - \frac{16 e^{-\f}}{C} (\p_8 \f - i\, \p_9 \f) \nonumber\\
&- 10 e^{-\f} \d^{mn} \p_m\f \p_n\f + \frac{16e^{-\f}}{C} (\p_8\f - i\, \p_9\f) = 0.
\end{align}

All other field equations have been checked and involve many
complicated cancellations. Carrying out these checks one obtains a
healthy respect for the nontriviality of this duality from the point
of view of the supergravity equations of motion.

\section{pp-wave}

Another type IIB background that is interesting to consider is the
pp-wave solution~\cite{hep-th/0110242}. This is a maximally supersymmetric
solution, and so by dualizing it with respect to any of its Killing
spinors we can get another maximally supersymmetric background of (complexified) type IIB supergravity.

In our conventions the pp-wave background is given by
\begin{subequations}
\be
ds^2 = 2 dx^+ dx^- - \l^2 \d_{\m\n} x^\m x^\n dx^+ dx^+ + \d_{\m\n} dx^\m dx^\n,
\ee
\be
\label{ppwave-RR}
F_{+1234} = 4\l = F_{+5678}
\ee
\end{subequations}
(in this section we use the lightcone coordinates $x^\pm =
\frac{1}{\sqrt{2}} (x^9 \pm x^0)$,
and $x^\m = \{x^1,\ldots,x^8\}$).
This solves the field equations for any constant $\l$:
the dilaton equation is $R=0$, which holds for the above metric, and
the only nontrivial Einstein equation is $R_{++} = \frac14
T^{(5)}_{++}$, which also holds with $R_{++} = 8\l^2$. All the other
equations are trivial due to the vanishing of almost all of the type IIB fields.

The Killing spinors of this background have been derived in~\cite{hep-th/0110242} and in our notation are given by
\be
\label{pp-killing}
\e = \left(\mathbb{1} - i x^\m \mathbb{A}_\m\right) \left(\cos\frac{\l x^+}{2} \mathbb{1} - i \sin\frac{\l x^+}{2}\mathbb{I}\right) \left(\cos\frac{\l x^+}{2} \mathbb{1} - i \sin\frac{\l x^+}{2}\mathbb{J}\right) \e_0,
\ee
for an arbitrary $\e_0$, where $\mathbb{1}$ is a $32 \times 32$ unit matrix, $\mathbb{I} = \G_1\G_2\G_3\G_4$, $\mathbb{J} = \G_5\G_6\G_7\G_8$, and
\be
\mathbb{A}_\m = \left\{
\begin{array}{ll}
8\l\, \G_-\, \mathbb{I}\, \G_\m, & \m = 1,2,3,4,\\
8\l\, \G_-\, \mathbb{J}\, \G_\m, & \m = 5,6,7,8.
\end{array}
\right.
\ee
The formula (\ref{pp-killing}) is written in the complex notation for the supersymmetry tranformations, see appendix~\ref{gamma}. Both $\e$ and $\e_0$ are Weyl spinors, i.e. complex, 16-component. Since full 32 by 32 gamma-matrices $\G_\m$ are used here, half of the components of $\e$ and $\e_0$ are zero.

In order to get the 32 basis elements $\{e_k = (\e_k,\eh_k)\}$ we first substitute arbitrary complex constants as the components of $\e_0$:
\be
\label{ab}
(\e_0)_k = \a_k + i \b_k,\qquad k\in\{1,\ldots,16\},\quad \a_k,\b_k \in \mathbb{R},
\ee
the rest 16 components of $\e_0$ being zero. Next we evaluate (\ref{pp-killing}) and get 16 complex components of $\e$.
Now, the real and imaginary parts of this Weyl spinor are our Killing spinors $(\e,\eh)$ in real notation. There are 32 independent pairs $e=(\e,\eh)$, corresponding to the thirty-two real parameters $\a_k$, $\b_k$.

The basis Killing spinor pairs then fall into two groups,
those that depend on $x^+$ only (`group $A$'),
and those that depend on the transverse coordinates $x^1,\ldots,x^8$
(`group $B$'). We get 16 group $A$ Killing spinors by keeping any of $\a_1,\ldots,\a_8$
(which we refer to as `group $A1$') or $\b_1,\ldots,\b_8$ (`group $A2$'),
while setting all other parameters to zero. Spinors that comprise group $B$
result from keeping any of $\a_9,\ldots,\a_{16}$ (`group $B1$') or
$\b_9,\ldots,\b_{16}$ (`group $B2$').

Not all of these Killing spinors satisfy the constraint (\ref{BM:constr}) (or its generalisation
(\ref{abel}), if one wants to perform multiple fermionic T-dualities).
If we pick a pair to construct a complex linear combination $f = e_a
+ i e_b$  so that $e_a$ and $e_b$ belong to different groups ($A$ and $B$),
then the condition (\ref{BM:constr}) cannot be satisfied.
Thus, necessarily $e_a, e_b\in A$ or $e_a,e_b\in B$.
According to the subdivision into subgroups $A1,A2,B1$, and $B2$,
there are four quite distinct fermionic T-dual backgrounds:
\begin{itemize}
\item $e_a,e_b \in A1$ or $e_a,e_b \in A2$;
\item $e_a,e_b \in B1$ or $e_a,e_b \in B2$;
\item $e_a\in B1$, $e_b \in B2$, or the other way round;
\item $e_a\in A1$, $e_b \in A2$, or the other way round.
\end{itemize}

The first case is much like the `simple' case of the transformed
D1-brane discussed in the section~\ref{easy} above. Namely, $C$ is
just a constant, dilaton is shifted by its logarithm and RR 5-form is
scaled by its power. Twenty-four new RR field components appear, eight
in $F_3$ and sixteen more in $F_5$. These look much like those given
in (\ref{F_3new}) and (\ref{F_5new}) multiplied additionally by a sine
or a cosine of $2 \l x^+$. Crucially, these new RR fluxes do not contribute to the stress-energy, precisely as in a D-brane case.

In the second case the transformed background is more complex.
It also has constant $C$, and therefore a constant dilaton and a
constant scaling factor for the 5-form components.
New in this case is that there are four nonvanishing components of RR
1-form, thirty-two components of the 3-form and fifty-six components of
the 5-form. All of these look like $\mathrm{const}\cdot (x^\m + i x^\n)$
for some $\m,\n\in \{1,\ldots,8\}$. Again, their stress-energy vanishes, so that no modification of the Einstein equations occurs.

The third case is interesting, the defining equation for $C$ is
nontrivial. We can proceed however forgetting about the factors of $C$
in all the RR form components. Three points are characteristic of a
dual background in this case:
there is no 3-form,
but all the 1-form and the 5-form components are nonzero;
all of these are either first or (more often) second order polynomials
in the transverse coordinates; and
they have nonvanishing stress-energy. The Einstein equations are still
satisfied due to the nontrivial spacetime dependence of the dilaton, which is proportional to $\log C$.

We will look in detail at the fourth case. This can be also
characterized by nontrivial contribution of the new components to the
stress-energy tensor, and a spacetime-dependent dilaton.

\subsection{Transformed pp-wave}
\label{pp-1}

The linear combination of the Killing spinors that we will use is $f = e_1 + i e_9$, where $e_1$ is what results from keeping only $\a_1 = 1$ in (\ref{ab}) while setting all the other parameters to zero (so this is a group $A1$ element), and $e_9$ corresponds to $\b_1 = 1$ (group $A2$). Explicitly this has the following form:
\be
\label{pp-1-k-sp}
f =
    \left\{
        \begin{aligned}
            &\{\cos \l x^+,0,0,i\sin \l x^+,0,0,0,0,0,0,0,0,0,0,0,0\}\\
            &\{i\cos \l x^+,0,0,-\sin \l x^+,0,0,0,0,0,0,0,0,0,0,0,0\}
        \end{aligned}
    \right\},
\ee
where the first line is $\e$ and the second line is $\eh$. This Killing spinor manifestly satisfies the constraint (\ref{BM:constr}), since in this case $\eh = i\e$, and thus
\be
\e\g_\m\e + \eh\g_\m\eh = \e\g_\m\e - \e\g_\m\e \equiv 0.
\ee

The defining equation for $C$ (\ref{BM:C}) takes the form
\be
\p_+ C = 2\sqrt{2}i \cos 2\l x^+, \quad\Rightarrow\quad C = \frac{i\sqrt{2}}{\l} \sin 2\l x^+.
\ee
The dilaton now depends on $x^+$:
\be
\f' = \frac12 \log \left(\frac{i\sqrt{2}}{\l} \sin 2\l x^+\right).
\ee

The RR 5-form components that were nonzero in the original background (\ref{ppwave-RR}) take the values
\be
F_{+1234} = F_{+5678} = 3\l \left( \frac{i\sqrt{2}}{\l} \sin 2\l x^+ \right)^{-1/2}.
\ee
The transformed background also has nonzero RR 1-form
\be
F_+ = -\cos 2\l x^+ \left( \frac{i\sqrt{2}}{\l} \sin 2\l x^+ \right)^{-3/2}
\ee
and the following new components of the 5-form:
\bsub
\begin{align}
F_{+1256} = F_{+1368} = F_{+1458} = F_{+2367} &= F_{+2457} = F_{+3478} \\
&= -\l \left( \frac{i\sqrt{2}}{\l} \sin 2\l x^+ \right)^{-1/2};\nonumber\\
F_{+1236} = F_{+1245} = F_{+3678} = F_{+4578} &= \left( \frac{i\sqrt{2}}{\l} \sin 2\l x^+ \right)^{-3/2};\\
F_{+1348} = F_{+1568} = F_{+2347} = F_{+2567} &= -\left( \frac{i\sqrt{2}}{\l} \sin 2\l x^+ \right)^{-3/2};\\
F_{+1278} = F_{+1467} = F_{+2358} = F_{+3456} &= \cos 2\l x^+ \left( \frac{i\sqrt{2}}{\l} \sin 2\l x^+ \right)^{-3/2};\\
F_{+1357} = F_{+2468} &= -\cos 2\l x^+ \left( \frac{i\sqrt{2}}{\l} \sin 2\l x^+ \right)^{-3/2}.
\end{align}
\esub

The only nonvanishing component of the energy-momentum tensors of these RR fields is the $(++)$ component, and this is readily calculated to give
\bsub
\begin{align}
T_{++}^{(1)} &= \frac{i\l^3}{\sqrt{2}} \frac{\cos^2 2\l x^+}{\sin^3 2\l x^+},\\
T_{++}^{(5)} &= 15\sqrt{2}i\l^3 \frac{\cos^2 2\l x^+}{\sin^3 2\l x^+} - 8\sqrt{2} i\l^3 \frac{1}{\sin^3 2\l x^+}.
\end{align}
\esub
The combination that enters the Einstein equations (\ref{_G}) is
\be
\frac{e^{2\f}}{2} \left( T_{++}^{(1)} + \frac12 T_{++}^{(5)} \right) = -8\l^2 \frac{\cos^2 2\l x^+}{\sin^2 2\l x^+} + 4\l^2 \frac{1}{\sin^2 2\l x^+}.
\ee
Recalling that $R_{++} = 8\l^2$ and calculating the second derivative of the dilaton to be
\be
\nabla_+\nabla_+\f = \p_+\p_+\f = -2\l^2 \frac{1}{\sin^2 2\l x^+},
\ee
we see that the Einstein equation (\ref{_G}) is satisfied by the transformed background:
\be
8\l^2 - 4\l^2 \frac{1}{\sin^2 2\l x^+} + 8\l^2 \frac{\cos^2 2\l x^+}{\sin^2 2\l x^+} - 4\l^2 \frac{1}{\sin^2 2\l x^+} \equiv 0.
\ee
All the other field equations are satisfied trivially.

\subsection{Purely imaginary fermionic T-dual background}

In the previous sections the transformed solutions were all
complex. Here we give an example of a solution that one can potentially make sense of within non-complexified supergravity. This is produced
by carrying out two independent fermionic T-dualities on the pp-wave. The result of the transformation has purely imaginary RR forms, so that timelike bosonic T-duality~\cite{hulltime} will make it real.

We begin by picking a second Killing spinor alongside with the one that has been used in the previous subsection:
\bsub
\label{pp-2}
\begin{align}
f_1 &=
    \left\{
        \begin{aligned}
            &\{\cos \l x^+,0,0,i\sin \l x^+,0,0,0,0,0,0,0,0,0,0,0,0\}\\
            &\{i\cos \l x^+,0,0,-\sin \l x^+,0,0,0,0,0,0,0,0,0,0,0,0\}
        \end{aligned}
    \right\},\\
f_2 &=
    \left\{
        \begin{aligned}
            &\{i\sin \l x^+,0,0,\cos \l x^+,0,0,0,0,0,0,0,0,0,0,0,0\}\\
            &\{-\sin \l x^+,0,0,i\cos \l x^+,0,0,0,0,0,0,0,0,0,0,0,0\}
        \end{aligned}
    \right\}.
\end{align}
\esub
The additional Killing spinor is a sum $f_2 = e_4 + i e_{12}$, where $e_4$ is a group $A1$ Killing spinor defined by $\a_4 = 1$ in (\ref{ab}) while setting all the other parameters to zero, and $e_{12}$ corresponds to $\b_4 = 1$ (group $A2$). The pair $(f_1,f_2)$ can be checked to satisfy (\ref{abel}).

The auxilliary function $C$ is a two by two matrix, defined by (\ref{C-mult}):
\be
C_{ij} =
    \left(
        \begin{array}{cc}
            a   & b\\
            b & a\\
        \end{array}
    \right),
\ee
where
\begin{align}
a &= \frac{i\sqrt{2}}{\l} \sin 2\l x^+,\\
b &= \frac{\sqrt{2}}{\l} \cos 2\l x^+.
\end{align}

The matrices $\log C$ and $C^{-1}$, which are needed in order to implement the formulae (\ref{mult-T-d}), have the same structure, but with different values for $a$ and $b$. Namely, we have for the inverse of $C$
\begin{align}
\label{a'}a' &= -\frac{i\l}{\sqrt{2}} \sin 2\l x^+,\\
\label{b'}b' &= \frac{\l}{\sqrt{2}} \cos 2\l x^+,
\end{align}
and for $\log C$:
\begin{align}
\label{a''}a'' &= \frac{i\pi}{2} + \log \frac{\sqrt{2}}{\l},\\
\label{b''}b'' &= -\frac{i\pi}{2} +  i\, 2\l x^+.
\end{align}

Using $\log C$ we can calculate the transformed dilaton:
\be
\f' = \frac12 \mathrm{Tr} \log C = a'' = \frac{i\pi}{2} + \log \frac{\sqrt{2}}{\l}, \qquad e^{\f'} = i\frac{\sqrt{2}}{\l}.
\ee
Thus the string coupling is purely imaginary in this background. From this we can already predict, that the transformed background will necessarily have purely imaginary RR flux, so that the sign of the combination $e^{2\f} F^2$ is invariant.

In order to derive this explicitly we calculate the contribution of the Killing spinors to the RR field strength bispinor, which is represented by the last term in (\ref{RR-mult}):
\be
\begin{aligned}
(\e_i \otimes \eh_j)\, (C^{-1})_{ij} =
                                            &-\frac{i\l}{\sqrt{2}} \sin 2\l x^+ \left[ \e_1 \otimes \eh_1 + \e_2 \otimes \eh_2 \right]\\
                                          &+ \frac{\l}{\sqrt{2}} \cos 2\l x^+ \left[\e_1 \otimes \eh_2 + \e_2 \otimes \eh_1 \right],
\end{aligned}
\ee
where $\e_i$ and $\eh_i$ are explicit components of $f_i = (\e_i,\eh_i)$. Substituting the values of the Killing spinors as given in (\ref{pp-2}), we arrive at the following background, which is indeed purely imaginary:
\bsub
\begin{align}
F_{+1234} &= F_{+5678} = -i \l^2 \sqrt{2} \\
F_{+1256} &= F_{+1368} = F_{+1458} = F_{+2367} = F_{+2457} = F_{+3478} = i \l^2 \sqrt{2}.
\end{align}
\esub
All other components of RR forms vanish. This background clearly satisfies Einstein equations, because
\be
R_{++} + 2\nabla_+\nabla_+ \f' - \frac{e^{2\f'}}{4} T^{(5)}_{++} = 8\l^2 - \frac14 \left(-\frac{2}{\l^2}\right) (-16 \l^4) \equiv 0.
\ee

\subsection{Self-duality of pp-wave}

We shall now show that the pp-wave background is self-dual under the
fermionic T-duality with respect to eight supersymmetries that we
denote by $\{f_1,\ldots,f_8\}$. Corresponding Killing spinors are all
of the same form as those
used to demonstrate how a single or double T-duality is done in the two previous subsections.
Namely, recapitulating the discussion after (\ref{ab}),
we pick sixteen real Killing spinors $\{e_1,\ldots,e_8\}\in A1$, $\{e_9,\ldots,e_{16}\} \in A2$.
Then the eight complex Killing spinors, satisfying (\ref{abel}), are given by
\be
f_i = e_i + i e_{i+8},\qquad i\in\{1,\ldots,8\}.
\ee
In particular, $f_1$ is exactly the same as $f$ that was used in section
\ref{pp-1} and was given by (\ref{pp-1-k-sp}).

With this choice of supersymmetries we get the following matrix $C$:
\be
C =
    \left(
        \begin{array}{cc}
            \begin{array}{cccc}
                a & 0 & 0 & b\\
                0 & a & -b& 0\\
                0 & -b& a & 0\\
                b & 0 & 0 & a\\
            \end{array}
            & 0\\
            0 &
            \begin{array}{cccc}
                a & 0 & 0 &-b\\
                0 & a & b & 0\\
                0 & b & a & 0\\
                -b& 0 & 0 & a\\
            \end{array}
        \end{array}
    \right),
\ee
where $a$ and $b$ are the same as in the previous subsection:
\begin{align}
a &= \frac{i\sqrt{2}}{\l} \sin 2\l x^+,\\
b &= \frac{\sqrt{2}}{\l} \cos 2\l x^+.
\end{align}

The matrices $\log C$ and $C^{-1}$ again have the same structure, but with different values for $a$ and $b$, which coincide with those given in the previous subsection, see eqs. (\ref{a'}) to (\ref{b''}).

The transformed dilaton is then evaluated to be
\be
\label{mult-T-d:new-dil}
\f' = 4 a'' = 2\pi i +4 \log \frac{\sqrt{2}}{\l}, \qquad e^{\f'} = \frac{4}{\l^4},
\ee
and
\be
\begin{aligned}
(\e_i \otimes \eh_j)\, (C^{-1})_{ij} =
                                            &-\frac{i\l}{\sqrt{2}} \sin 2\l x^+ \left[ \e_1 \otimes \eh_1 + \ldots + \e_8 \otimes \eh_8 \right]\\
                                          &+ \frac{\l}{\sqrt{2}} \cos 2\l x^+
                                                    \left[
                                                        \begin{aligned}
                                                            &\e_1 \otimes \eh_4 + \e_4 \otimes \eh_1 - \e_2 \otimes \eh_3 - \e_3 \otimes \eh_2\\
                                                            -&\e_5 \otimes \eh_8 - \e_8 \otimes \eh_5 + \e_6 \otimes \eh_7 + \e_7 \otimes \eh_6
                                                        \end{aligned}
                                                    \right],
\end{aligned}
\ee
where $(\e_i,\eh_i)=f_i$.

An important feature of this matrix, which becomes obvious only after explicit substitution of the Killing spinors, is that it is proportional to the first term on the right-hand side of (\ref{RR-mult}). This leads to the RR field bispinor after the transformation being proportional to itself before the transformation. More precisely, we have for the transformed RR background
\be
F_{+1234} = -\l^5 = F_{+5678},
\ee
with all other components vanishing. This is just the original flux that was supporting the pp-wave geometry before we have done fermionic T-duality, but multiplied by a constant $-\frac{\l^4}{4}$. Since this constant is equal to $-e^{-\f'}$ (\ref{mult-T-d:new-dil}), the Einstein equations hold for the new background because they involve a product $e^{2\f'} T_{\m\n}^{(5)}$:
\be
R_{++} + 2\nabla_+\nabla_+ \f' - \frac{e^{2\f'}}{4} T^{(5)}_{++} = 8\l^2 - \frac14 \left(\frac{4}{\l^4}\right)^2 (\l^{10} + \l^{10}) \equiv 0.
\ee

This transformation clearly leaves the string spectrum invariant since it is just a field redefinition of the Ramond-Ramond field strength. 

Interestingly, if one splits the eight supersymmetries that were used in this section into two groups $\{f_1,\ldots,f_4\}$ and $\{f_5,\ldots,f_8\}$ and performs fermionic T-dualities of the original pp-wave background with respect to each of these groups independently, then the resulting background has the dilaton $e^{\f'} = \frac{2}{\l^2}$ in both cases, and the RR forms in the two cases are given by
\be
F_{+1458} = F_{+2367} = \pm 2\l^3.
\ee
Thus each group of four fermionic T-dualities also results in a pp-wave background that has undergone a certain rotation in transverse directions as compared to the original pp-wave.

\section{Discussion}

Fermionic T-duality has many interesting properties, some of them are
quite unexpected. First, one should note that fermionic T-duality does not commute with bosonic T-duality. This is easily seen with the D1-brane case where new Ramond-Ramond fields are produced that break the $\mathrm{SO}(1,1) \times \mathrm{SO}(8)$ symmetry of the original D1-brane solution. In retrospect this should not be a surprise since it is known that supersymmetries and isometries do not commute either. One can also think of examples where T-duality breaks supersymmetry (at the level of supergravity).

We have also checked to see whether fermionic T-duality is nilpotent and we see that it is not always so. In the examples carried out above the transformation is only nilpotent up to a root of unity. This is undoubtedly a consquence of the fermionic nature of the transformation.

One of the main goals of this paper was to find transformations to real solutions. This has been successful in that we have shown that the pp-wave can be transformed to produce real solutions but in that case the transformed solution is again the pp-wave up to some field redefinitions or rotations. 

It seems somewhat distant at this point to be able to know when a real solution is possible and what the new solutions will be. We intend to pursue this question in futher work.

\section{Acknowledgements}

We are grateful to Nathan Berkovits, Matthias Blau, Nick Dorey, Giuseppe
Policastro, Daniel Thompson and Gabriele Travaglini for clarifying
various points. In particular DSB
wishes to thank Malcolm Perry for extensive discussions at the
beginning of this project and DAMTP Cambridge for continued hospitality.
IB is supported by a Westfield Trust Scholarship. DSB is partly
funded by the STFC rolling grant.

\appendix

\section{Type IIB supergravity action and equations of motion}
\label{app-sugra}

We will give the relevant action and equations of motion for IIB
supergravity so that all our conventions are transparent. Our metric signature is mostly plus, $(-\,+\ldots +)$; antisymmetric Levi-Civita tensor is defined with $\e_{0\ldots 9} = 1$.
Apart from the metric, which is represented by $g_{\m\n}$, the bosonic
field content of type IIB supergravity is given by two real scalars,
dilaton $\f$ and RR scalar $C_0$, two real
antisymmetric second-rank tensors $B$ and $C_2$ and a fourth-rank real
tensor $C_4$, whose field stregth $F_5=dC_4$ is self-dual:
\be
F_{\m_1\ldots\m_5} =  \frac{1}{5!} \e_{\m_1\ldots\m_5\n_1\ldots\n_5} F^{\n_1\ldots\n_5}.
\ee
From string theory point of view, the fields $C_0, C_2$, and $C_4$ are
potentials of the RR fields $F_{n+1} = dC_n$. Three remaining fields
$g, B$, and $\f$ belong to the NSNS sector of type IIB superstring.

The action of type IIB supergravity in the string frame is a sum of three terms
\be
\label{action}
S = S_{NSNS} + S_{RR} + S_{CS},
\ee
where
\begin{align}
S_{NSNS} &= \frac{1}{2\k^2}\int d^{10}x \sqrt{|g|} \,e^{-2\f} \left[R+4(\p\f)^2-\frac12\frac{1}{3!}{H_3}^2\right],\\
S_{RR} &= -\frac{1}{4\k^2}\int d^{10}x \sqrt{|g|} \left[{F_1}^2 + \frac{1}{3!}\tilde {F_3}^2 +  \frac12\frac{1}{5!} \tilde {F_5}^2 \right],\\
S_{CS} &= - \frac{1}{4\k^2}\int C_4\wedge H_3\wedge F_3.
\end{align}
Here $H_3 = dB_2$ is the field strength of the NSNS antisymmetric tensor field, and we use a common notation ${F_n}^2 = F_{\m_1\ldots \m_n} F_{\n_1\ldots \n_n} g^{\m_1 \n_1} \ldots g^{\m_n \n_n}$. Modified field strengths $\tilde F_n$ are used in $S_{RR}$, and only there:
\bsub
\label{mod}
\begin{align}
\tilde F_3 &= F_3 - C_0 H_3,\\
\tilde F_5 &= F_5 - \frac12 C_2 \wedge H_3 + \frac12 B_2 \wedge F_3.
\end{align}
\esub
Note that these reduce to ordinary $F_n$ if the $B$-field is zero.

The equations of motion of the two scalars in the
theory (\ref{action}) are the simplest. The dilaton equation reads
\be
\label{phi}
R = 4(\p\f)^2 - 4\nabla^2 \f + \frac12\frac{{H_3}^2}{2},
\ee
and the RR scalar field equation is
\be
\label{C_0}
\nabla^2 C_0 + \frac{1}{3!} H_3 \tilde F_3 = 0.
\ee

The equations for $B_2, C_2$, and $C_4$ are respectively (note that the first two equations have been simplified somewhat by substitution of the third one):
\begin{align}
\label{B_2}
\nabla_\m&\left[e^{-2\f} H - C_0 \tilde F\right]^{\a\b\m} \nonumber\\
& = \frac12\frac{1}{3!} \tilde F^{\a\b\m\n\l} F_{\m\n\l} - \frac{1}{2\sqrt{|g|}}\frac{1}{5!}\frac{1}{3!} \e^{\a\b\m_1\ldots \m_5 \n_1\ldots \n_3} \tilde F_{\m_1\ldots \m_5} F_{\n_1\ldots \n_3};
\end{align}
\begin{align}
\label{C_2}
\nabla_\m &\tilde F^{\a\b\m} \nonumber\\
& = - \frac12\frac{1}{3!} \tilde F^{\a\b\m\n\l} H_{\m\n\l} + \frac{1}{2\sqrt{|g|}}\frac{1}{5!}\frac{1}{3!} \e^{\a\b\m_1\ldots \m_5 \n_1\ldots \n_3} \tilde F_{\m_1\ldots \m_5} H_{\n_1\ldots \n_3};
\end{align}
\be
\label{C_4}
\nabla_\m \tilde F^{\m\n_1\ldots \n_4} = \frac{1}{\sqrt{|g|}}\frac{1}{3!}\frac{1}{3!}\e^{\n_1\ldots \n_4 \l_1\ldots \l_3 \r_1\ldots \r_3} H_{\l_1\ldots \l_3} F_{\r_1\ldots \r_3}.
\ee

Finally the Einstein equations, after simplifying by substitution
of the Ricci scalar as given by the dilaton equation (\ref{phi}) are:
\be
\label{G}
R_{\m\n} + 2\nabla_\m\nabla_\n\f = \frac14 H_{\m\a\b} {H_\n}^{\a\b} + \frac{e^{2\f}}{2} \left[ T_{\m\n}^{(1)} + T_{\m\n}^{(\tilde 3)} + \frac12 T_{\m\n}^{(\tilde 5)} \right],
\ee
where
\begin{align}
T_{\m\n}^{(1)} &= \p_\m C \p_\n C - \frac12 g_{\m\n} (\p C)^2,\\
T_{\m\n}^{(\tilde 3)} &= \frac12 \tilde F_{\m\a\b}  {\tilde F_\n}^{~\,\a\b} - \frac12 g_{\m\n} \frac{1}{3!}\tilde {F_3}^2 ,\\
T_{\m\n}^{(\tilde 5)} &= \frac{1}{4!} \tilde F_{\m\a_1\ldots \a_4}  {\tilde F_\n}^{~\,\a_1\ldots \a_4}
\end{align}
(the $\tilde {F_5}^2$ term in the 5-form energy-momentum is identically zero since $\tilde F_5 = \star \tilde F_5$).

The supergravity field equations, which we have derived here, simplify
considerably in the case of zero $B$-field, as is relevent for D-brane
solutions. For the dilaton, RR scalar, $B_2$, $C_2$, $C_4$, and $g$ we have correspondingly
\begin{align}
\label{_phi} &R = 4(\p\f)^2 - 4\nabla^2 \f,\\
\label{_C_0} &\nabla^2 C_0 = 0,\\
\label{_B_2} &\nabla_\m\left(C_0 F\right)^{\a\b\m} = -\frac12\frac{1}{3!} F^{\a\b\m\n\l} F_{\m\n\l}+ \frac{1}{2\sqrt{|g|}}\frac{1}{5!}\frac{1}{3!} \e^{\a\b\m_1\ldots \m_5 \n_1\ldots \n_3} F_{\m_1\ldots \m_5} F_{\n_1\ldots \n_3},\\
\label{_C_2} &\nabla_\m F^{\a\b\m} = 0,\\
\label{_C_4} &\nabla_\m F^{\m\n_1\ldots \n_4} = 0,\\
\label{_G} &R_{\m\n} + 2\nabla_\m\nabla_\n\f = \frac{e^{2\f}}{2} \left[ T_{\m\n}^{(1)} + T_{\m\n}^{(3)} + \frac12 T_{\m\n}^{(5)} \right].
\end{align}

\section{Gamma-matrices and supersymmetry transformations}
\label{gamma}

We work with the real 32 by 32 representation for the gamma-matrices of $(9+1)$-dimensional spacetime, that exist due to the isomorphism $\Cl(9,1) \cong \mathrm{Mat}(\mathbb{R},32)$. It is convenient to exploit the periodicity property of the Clifford algebras
\be
\Cl(9,1) \cong \Cl(1,1) \otimes \Cl(8,0)
\ee
to construct the gamma-matrices as tensor products of $\{\s_1, i\s_2\}$, which are the gamma-matrices of $Cl(1,1)$ with the following symmetric $\{\S_1,\ldots,\S_8\}$, which are the gamma-matrices of $8$-dimensional Euclidean space:
\be
\begin{array}{ccccccc}
\S^1 = \sigma_2 & \otimes & \sigma_2 & \otimes & \sigma_2 & \otimes & \sigma_2, \\
\S^2 = \sigma_2 & \otimes & 1             & \otimes & \sigma_1 & \otimes & \sigma_2, \\
\S^3 = \sigma_2 & \otimes & 1             & \otimes & \sigma_3 & \otimes & \sigma_2, \\
\S^4 = \sigma_2 & \otimes & \sigma_1 & \otimes & \sigma_2 & \otimes & 1,\\
\S^5 = \sigma_2 & \otimes & \sigma_3 & \otimes & \sigma_2 & \otimes & 1,\\
\S^6 = \sigma_2 & \otimes & \sigma_2 & \otimes & 1             & \otimes & \sigma_1, \\
\S^7 = \sigma_2 & \otimes & \sigma_2 & \otimes & 1             & \otimes & \sigma_3, \\
\S^8 = \sigma_1 & \otimes & 1             & \otimes & 1              & \otimes & 1,
\end{array}
\ee
and $\S^9=\S^1\cdot\ldots\cdot\S^8 = \s_3 \otimes  1 \otimes 1 \otimes 1$, which is a chirality operator in 8D.
In particular, the representation we use is:
\be
\begin{aligned}
\G^0 = i\sigma_2 \otimes \mathbb{1}_{16} & = \left( \begin{array}{cc}
                                                                                    0 & \mathbb{1}_{16}\\ -\mathbb{1}_{16} & 0
                                                                                    \end{array}\right), \qquad (\G^0)^2 = -1;\\
\G^i = \sigma_1 \otimes \S^i & = \left( \begin{array}{cc}
                                                                0 & \S^i \\ \S^i & 0
                                                            \end{array}\right), \qquad (\G^i)^2 = 1.
\end{aligned}
\ee

The 10-dimensional chirality operator is $\G^{10} = \G^0\cdot\ldots\cdot\G^9 = \s_3\otimes\mathbb{1}_{16}$. Spinors of definite chirality are defined as usual, $\G^{10}\psi^\pm = \pm\psi^\pm$; they provide two inequivalent real 16-dimensional representations of $\Spin(9,1)$, $S_+$ and $S_-$. These are Majorana-Weyl spinors; we can also define $S_+ \oplus S_-$, which is a Majorana spinor (real 32 component) and $S_+ \otimes \mathbb{C}$ ($S_- \otimes \mathbb{C}$), which are Weyl spinors (complex 16 component) of positive (negative) chirality.

The $\g^\m$ matrices, which are used throughout the paper, are defined as off-diagonal 16 by 16 blocks of the $\G^\m$ matrices:
\be
\begin{aligned}
\G^\m =
    \left(
        \begin{array}{cc}
            0 & {\g^\m}^{\a\b}\\
            \g^\m_{\a\b} & 0\\
        \end{array}
    \right),
\end{aligned}
\ee
so that they are analogs of Pauli matrices in 4D. Explicitly
\bsub
\begin{align}
{\g^\m}^{\a\b} &= (1,\S^i),\\
\g^\m_{\a\b} &= (-1,\S^i).
\end{align}
\esub
The $\g^\m$ matrices are symmetric and they satisfy a condition
\be
\g^\m_{\a\b} {\g^\n}^{\b\g} + \g^\n_{\a\b} {\g^\m}^{\b\g} = 2 \eta^{\m\n} \d_\a^\g.
\ee
Position of the spinor indices reflects the convention to denote the positive chirality spinors with $\psi^\a$ and the negative chirality spinors with $\chi_\a$. For example, action of a gamma-matrix on a Majorana spinor is given by
\be
\G^\m \Psi =
    \left(
        \begin{array}{cc}
            0 & {\g^\m}^{\a\b}\\
            \g^\m_{\a\b} & 0\\
        \end{array}
    \right)
    \left(
        \begin{array}{c}
            \psi^\b \\
            \chi_\b\\
        \end{array}
    \right) =
\left(
        \begin{array}{c}
            (\g^\m \chi)^\a\\
            (\g^\m \psi)_\a\\
        \end{array}
    \right),
\ee
and action on chiral (Majorana-Weyl or Weyl) spinors can be written by setting $\psi$ or $\chi$ to zero.

Since a charge conjugation matrix in this representation can be taken to be $C=\G^0$:
\be
C \G^i C^{-1} = -{\G^i}^T,
\ee
the Lorentz-covariant bilinear takes the form (using Majorana conjugation $\overline\Psi = \Psi^T C$):
\be
\begin{aligned}
\overline\Psi \G^\m \Phi &=
    \left(
        \begin{array}{cc}
            \psi^\a &   \chi_\a\\
        \end{array}
    \right)
    \left(
        \begin{array}{cc}
            0 & {1_\a}^\b\\
            -{1^\a}_\b & 0\\
        \end{array}
    \right)
    \left(
        \begin{array}{cc}
            0 & {\g^\m}^{\b\g}\\
            \g^\m_{\b\g} & 0\\
        \end{array}
    \right)
    \left(
        \begin{array}{c}
            \phi^\g \\
            \varphi_\g\\
        \end{array}
    \right) =\\
&=\psi^\a \g^\m_{\a\b} \phi^\b - \chi_\a {\g^\m}^{\a\b} \varphi_\b.
\end{aligned}
\ee
For chiral spinors, such as the supersymmetry parameters of IIB supergravity, this bilinear reduces to $\psi^\a\g^\m_{\a\b} \phi^\b$ (in the case of positive chirality). This type of 16-component spinor bilinear is used, e.g. in the formula (\ref{BM:constr}).

Killing spinor equations result from requiring that the supersymmetry variations of the fermions vanish. The fermions in type IIB supergravity are the doublets of gravitini and dilatini, which have opposite chirality. We take the dilatini $\l,\hat\l$ to have negative chirality. The supersymmetry parameters $\e,\eh$ are of the same (positive) chirality as the gravitini $\psi_\m,\hat\psi_\m$. Supersymmetry variations in the two-component formalism are:
\begin{align}
\d\psi_\m &= \nabla_\m\e - \frac14 \slashed{H}_\m\e - \frac{e^\f}{8} \left( \slashed{F}_1 + \slashed{F}_3 + \frac12 \slashed{F}_5 \right) \G_\m \eh, \\
\d\hat\psi_\m &= \nabla_\m\eh + \frac14 \slashed{H}_\m \eh + \frac{e^\f}{8} \left( \slashed{F}_1 - \slashed{F}_3 + \frac12 \slashed{F}_5 \right) \G_\m \e,
\end{align}
\begin{align}
\d\l_\m &= \slashed{\p}\f\e -\frac12 \slashed{H} \e + \frac{e^\f}{2} \left( 2\slashed{F}_1 + \slashed{F}_3 \right) \eh,\\
\d\hat\l_\m &= \slashed{\p}\f\eh +\frac12 \slashed{H} \eh - \frac{e^\f}{2} \left( 2\slashed{F}_1 - \slashed{F}_3 \right) \e,
\end{align}
where
\begin{align}
\slashed{F}_n &= \frac{1}{n!} F_{\m_1\ldots\m_n} \G^{\m_1\ldots\m_n},\\
\slashed{H}_\m &= \frac12 H_{\m\n\r} \G^{\n\r}.
\end{align}
Sometimes it is more convenient to derive and solve the Killing spinor equations in terms of the single complex gravitino, dilatino and supersymmetry parameter, defined as
\be
\Psi_\m = \psi_m + i \hat\psi_\m, \quad\Lambda = \l + i \hat\l, \quad\varepsilon = \e + i\eh.
\ee
The above transformations can be rewritten in the complex notation as
\be
\d\Psi_\m = \nabla_\m \varepsilon -\frac14 \slashed{H}_\m \varepsilon^* + \frac{i e^\f}{8} \left(  \slashed{F}_1 + \frac12 \slashed{F}_5 \right) \G_\m \varepsilon - \frac{i e^\f}{8} \slashed{F}_3 \G_\m\varepsilon^*,
\ee
\be
\d\Lambda =  \slashed{\p}\f\varepsilon -\frac12 \slashed{H} \varepsilon^* -i e^\f \slashed{F}_1 \varepsilon + \frac{i e^\f}{2} \slashed{F}_3 \varepsilon^*.
\ee


\begin{thebibliography}{1}




%\cite{Giveon:1994fu}
\bibitem{eliezer}
  A.~Giveon, M.~Porrati and E.~Rabinovici,
  ``Target space duality in string theory,''
  Phys.\ Rept.\  {\bf 244} (1994) 77
  [arXiv:hep-th/9401139].
  %%CITATION = PRPLC,244,77;%%






%\cite{Buscher:1987qj}
\bibitem{Buscher}
  T.~H.~Buscher,
  ``Path Integral Derivation of Quantum Duality in Nonlinear Sigma Models,''
  Phys.\ Lett.\  B {\bf 201} (1988) 466.
  %%CITATION = PHLTA,B201,466;%%





%\cite{Bergshoeff:1995as}
\bibitem{Berg}
  E.~Bergshoeff, C.~M.~Hull and T.~Ortin,
  ``Duality in the type II superstring effective action,''
  Nucl.\ Phys.\  B {\bf 451}, 547 (1995)
  [arXiv:hep-th/9504081].
  %%CITATION = NUPHA,B451,547;%%


%\cite{Maldacena:1999mh}
\bibitem{maldanc}
  J.~M.~Maldacena and J.~G.~Russo,
  ``Large N limit of non-commutative gauge theories,''
  JHEP {\bf 9909}, 025 (1999)
  [arXiv:hep-th/9908134].
  %%CITATION = JHEPA,9909,025;%%
  M.~Alishahiha, Y.~Oz and M.~M.~Sheikh-Jabbari,
  ``Supergravity and Large N Noncommutative Field Theories,''
  JHEP {\bf 9911}, 007 (1999)
  [arXiv:hep-th/9909215].
  %%CITATION = JHEPA,9911,007;%%
R.~G.~Cai and N.~Ohta,
  ``On the thermodynamics of large N non-commutative super Yang-Mills
  theory,''
  Phys.\ Rev.\  D {\bf 61} (2000) 124012
  [arXiv:hep-th/9910092],





%\cite{Lunin:2005jy}
\bibitem{lunin}
  O.~Lunin and J.~M.~Maldacena,
  ``Deforming field theories with U(1) x U(1) global symmetry and their
  gravity duals,''
  JHEP {\bf 0505}, 033 (2005)
  [arXiv:hep-th/0502086].
  %%CITATION = JHEPA,0505,033;%%

%\cite{Gursoy:2005cn}
\bibitem{dipole}
  U.~Gursoy and C.~Nunez,
  ``Dipole deformations of N = 1 SYM and supergravity backgrounds with U(1)  x
  U(1) global symmetry,''
  Nucl.\ Phys.\  B {\bf 725}, 45 (2005)
  [arXiv:hep-th/0505100].
  %%CITATION = NUPHA,B725,45;%%

%\cite{Berman:2007tf}
\bibitem{laura}
  D.~S.~Berman and L.~C.~Tadrowski,
  ``M-Theory Brane Deformations,''
  Nucl.\ Phys.\  B {\bf 795}, 201 (2008)
  [arXiv:0709.3059 [hep-th]].
  %%CITATION = NUPHA,B795,201;%%





%\cite{Berkovits:2008ic}
\bibitem{BerkMald}
  N.~Berkovits and J.~Maldacena,
  ``Fermionic T-Duality, Dual Superconformal Symmetry, and the Amplitude/Wilson
  Loop Connection,''
  JHEP {\bf 0809}, 062 (2008)
  [arXiv:0807.3196 [hep-th]].
  %%CITATION = JHEPA,0809,062;%%


\bibitem{Beisert:2008iq}
  N.~Beisert, R.~Ricci, A.~A.~Tseytlin and M.~Wolf,
  %``Dual Superconformal Symmetry from AdS5 x S5 Superstring Integrability,''
  Phys.\ Rev.\  D {\bf 78} (2008) 126004
  [arXiv:0807.3228 [hep-th]].
  %%CITATION = PHRVA,D78,126004;%%



%\cite{Benichou:2008it}
\bibitem{policastro}
  R.~Benichou, G.~Policastro and J.~Troost,
  %``T-duality in Ramond-Ramond backgrounds,''
  Phys.\ Lett.\  B {\bf 661}, 192 (2008)
  [arXiv:0801.1785 [hep-th]].
  %%CITATION = PHLTA,B661,192;%%



%\cite{Hull:1998ym}
\bibitem{hulltime}
  C.~M.~Hull,
  ``Duality and the Signature of Space-Time,''
  JHEP {\bf 9811} (1998) 017
  [arXiv:hep-th/9807127].
  %%CITATION = JHEPA,9811,017;%%
  C.~M.~Hull,
  ``Timelike T-duality, de Sitter space, large N gauge theories and
  topological field theory,''
  JHEP {\bf 9807} (1998) 021
  [arXiv:hep-th/9806146].
  %%CITATION = JHEPA,9807,021;%%



\bibitem{gab}
 R.~Britto, F.~Cachazo, B.~Feng and E.~Witten,
  ``Direct Proof Of Tree-Level Recursion Relation In Yang-Mills
Theory,''
  Phys.\ Rev.\ Lett.\  {\bf 94} (2005) 181602
  [arXiv:hep-th/0501052].
  %%CITATION = PRLTA,94,181602;%%
  L.~F.~Alday and J.~M.~Maldacena,
  %``Gluon scattering amplitudes at strong coupling,''
  JHEP {\bf 0706} (2007) 064
  [arXiv:0705.0303 [hep-th]].
  %%CITATION = JHEPA,0706,064;%%
  J.~M.~Drummond, G.~P.~Korchemsky and E.~Sokatchev,
  ``Conformal properties of four-gluon planar amplitudes and Wilson
loops,''
  Nucl.\ Phys.\  B {\bf 795} (2008) 385
  [arXiv:0707.0243 [hep-th]].
  %%CITATION = NUPHA,B795,385;%%
  A.~Brandhuber, P.~Heslop and G.~Travaglini,
  ``MHV Amplitudes in N=4 Super Yang-Mills and Wilson Loops,''
  Nucl.\ Phys.\  B {\bf 794} (2008) 231
  [arXiv:0707.1153 [hep-th]].
  %%CITATION = NUPHA,B794,231;%%



%\cite{Adam:2009kt}
\bibitem{oz}
  I.~Adam, A.~Dekel and Y.~Oz,
  ``On Integrable Backgrounds Self-dual under Fermionic T-duality,''
  JHEP {\bf 0904}, 120 (2009)
  [arXiv:0902.3805 [hep-th]].
  %%CITATION = JHEPA,0904,120;%%

%\cite{Fre:2009ki}
\bibitem{grassi}
  P.~Fre, P.~A.~Grassi, L.~Sommovigo and M.~Trigiante,
  ``Theory of Superdualities and the Orthosymplectic Supergroup,''
  Nucl.\ Phys.\  B {\bf 825}, 177 (2010)
  [arXiv:0906.2510 [hep-th]].
  %%CITATION = NUPHA,B825,177;%%

\bibitem{Chen}
C.~g.~Hao, B.~Chen and X.~c.~Song,
  ``On Fermionic T-duality of Sigma modes on AdS backgrounds,''
  JHEP12(2009)051, arXiv:0909.5485 [hep-th].





\bibitem{HorStrom}
  G.~Horowitz and A.~Strominger,
  ``Black strings and P-branes,''
  Nucl.\ Phys.\ B  {\bf 360} (1991) 197.


\bibitem{DKL}
  M.~J.~Duff, R.~R.~Khuri and J.~X.~Lu,
  ``String solitons,''
  Phys.\ Rept.\  {\bf 259} (1995) 213
  [arXiv:hep-th/9412184].
  %%CITATION = PRPLC,259,213;%%




\bibitem{hep-th/0110242}
M.~Blau, J.~M.~Figueroa-O'Farrill, C.~Hull and G.~Papadopoulos,
  ``A new maximally supersymmetric background of IIB superstring theory,''
  JHEP {\bf 0201} (2002) 047
  [arXiv:hep-th/0110242].
  %%CITATION = JHEPA,0201,047;%%

%\cite{Berenstein:2002jq}
\bibitem{BMN}
  D.~E.~Berenstein, J.~M.~Maldacena and H.~S.~Nastase,
  ``Strings in flat space and pp waves from N = 4 super Yang Mills,''
  JHEP {\bf 0204}, 013 (2002)
  [arXiv:hep-th/0202021].
  %%CITATION = JHEPA,0204,013;%%



\end{thebibliography}
\end{document}